\documentstyle[twoside,fleqn,espcrc2,epsfig]{article}
\def\etal{{\it et.al.}\/}
\def\ie{{\it i.e.}\/}

\def\fip{{\Phi_{\mbox{\footnotesize pp+pep}}}\/}
\def\fibe{{\Phi_{\mbox{\footnotesize Be}}}\/}
\def\fib{{\Phi_{\mbox{\footnotesize B}}}\/}
\def\ficno{{\Phi_{\mbox{\footnotesize CNO}}}\/}
\def\fibecno{{\Phi_{\mbox{\footnotesize Be+CNO}}}\/}
\title{INTERMEDIATE ENERGY SOLAR NEUTRINOS}
\author{E. Calabresu \address{Dipartimento di Fisica
        dell'Universit\`a di Cagliari, I-09124 Cagliari \\
        and Istituto Nazionale di Fisica Nucleare,
        Sezione di Cagliari, I-09127 Cagliari, Italy},
        G. Fiorentini\address{Dipartimento di Fisica
         dell'Universit\`a di Ferrara, I-44100 Ferrara \\
        and Istituto Nazionale di Fisica Nucleare, Sezione di Ferrara,
        I-44100 Ferrara, Italy},
         M.~Lissia$^{\mbox{\footnotesize a}}$
        and
         B. Ricci\address{Dipartimento di Fisica
         dell'Universit\`a di Padova, I-35100 Padova \\
        and Istituto Nazionale di Fisica Nucleare, Sezione di Ferrara,
        I-44100 Ferrara, Italy}
        }

\begin{document}

\begin{abstract}
We extract information on the fluxes of
Be and CNO neutrinos directly from solar neutrino experiments, with
minimal assumptions about solar models.
Next we compare these results with solar models, both standard
and non standard ones.
Finally we discuss the expectations for Borexino, both in the case of
standard and non standard neutrinos.
\end{abstract}

\maketitle

\section{Introduction}

The principal aim of this paper is to extract information on the fluxes of
Be and CNO neutrinos directly from solar neutrino experiments, with
minimal assumptions about solar models.
In this respect, we will update previous
results~\cite{AA,PRD,WHERE,HATA1,HATA2}
and try to elucidate the role of CNO
neutrinos. We will see that experimental data are more and more against
the hypothesis of standard neutrinos (\ie ~without mass, mixing, magnetic
moments...).

Next we will compare these informations with solar models, both standard
and non standard ones. Clearly, low (\ie ~smaller than standard) central
temperature models are ruled out, essentially because they  cannot
reproduce  the  experimental data available on both Be  and B neutrinos.
Hybrid models, where some suitable nuclear cross section is varied in
order to reduce the Be neutrinos flux to the observed value and  with a higher
central temperature, so as to  agree with experimental results on
B neutrinos flux,
can also be excluded, as in these models the CNO neutrino
flux grows beyond
acceptable levels. In other words, the  bounds on Be (CNO) neutrinos tell
us that it is hopeless to reduce (enhance) the central solar temperature,
in order to stay with standard neutrinos.

In summary, we shall demonstrate that, under the assumption of standard
neutrinos:
\begin{itemize}
\item the available experimental results look inconsistent among themselves,
even if one of the four experiments were wrong;

\item the flux of intermediate energy neutrinos (Be+CNO) as derived from
experiments is significantly smaller than the prediction of SSM's;

\item the different reduction factors for $^7$Be and $^8$B neutrinos
with  respect to the SSM are essentially in
contradiction with the fact that both $^7$Be and $^8$B neutrinos
originate from the same parent $^7$Be nucleus.
\end{itemize}

We will discuss then the expectations for Borexino, both in the case of
standard and non standard neutrinos, showing that the experiment can
clearly discriminate among several possible solutions to the solar
neutrino puzzle.
\begin{table*}[htb]
\setlength{\tabcolsep}{1.9pc}
\catcode`?=\active \def?{\kern\digitwidth}
\caption[pesi]{
For the $i$-th neutrino flux, we show the average neutrino energy
$\langle E\rangle _i$ and the energy averaged capture cross sections
in Chlorine ($\sigma_{i,C}$) and Gallium ($\sigma_{i,G}$).
Errors correspond to 1 standard deviation
and 1~SNU cm$^2$ s = 10$^{-36}$ cm$^2$.
 All data are from~\cite{libro}, but for $\sigma_{B,C}$
taken from~\cite{sfiga}.
When averaging the pp and pep components to get p, we use the relative
weights of the SSM from Ref.~\cite{BP95};
similarly for $^{13}$N and $^{15}$O to get CNO.
               }
\label{sigma}
\vspace{0.2cm}
\begin{tabular}{lr@{}lr@{}l@{}lr@{}l@{}l}
\hline
         &\multicolumn{2}{c}{$\langle E \rangle_i$}
                    &\multicolumn{3}{c}{$\sigma_{i,C}$}
                                      &\multicolumn{3}{c}{$\sigma_{i,G}$} \\
         &\multicolumn{2}{c}{[MeV]}
                    &\multicolumn{3}{c}{[10$^{-9}$SNU cm$^2$s]}
                         &\multicolumn{3}{c}{[10$^{-9}$SNU cm$^2$s]} \\
\hline
pp       & 0. & 265 &    0. &    &               &    1. & 18 & $(1\pm 0.02)$\\
pep      & 1. & 442 &    1. & 6  & $(1\pm 0.02)$ &   21. & 5  & $(1\pm 0.07)$\\
p=pp+pep & 0. & 268 &    0. & 38$\cdot10^{-2}$& $(1\pm 0.02)$
                                                 &    1. & 23 & $(1\pm 0.02)$\\
$^7$Be   & 0. & 814 &    0. & 24 & $(1\pm 0.02)$ &    7. & 32 & $(1\pm 0.03)$\\
$^{13}$N & 0. & 707 &    0. & 17 & $(1\pm 0.02)$ &    6. & 18 & $(1\pm 0.03)$\\
$^{15}$O & 0. & 996 &    0. & 68 & $(1\pm 0.02)$ &   11. & 6  & $(1\pm 0.06)$\\
CNO=$^{13}$N + $^{15}$O
         & 0. & 842 &    0. & 41 & $(1\pm 0.02)$ &    8. & 72 & $(1\pm 0.05)$\\
$^8$B    & 6. & 71  &    1. & 11$\cdot10^{3}$
                        & $(1\pm 0.03)$ & 2. & 43$\cdot10^{3}$   & $(1\pm
0.25)$\\
\hline
\end{tabular}
\end{table*}%
\section{Where are Be and CNO neutrinos ?}

We make the assumption of stationary Sun (\ie ~the presently observed
luminosity equals  the  present nuclear energy production rate) and
standard neutrinos, so that  all the  $\nu_e$ produced in the Sun reach Earth
without being lost and  their energy spectrum is unchanged.  The relevant
variables are thus the (energy integrated) neutrino fluxes, which can be
grouped as:
\begin{equation}
\fip, \quad \fibe, \quad \ficno \quad {\mbox{and}}\quad \fib \quad.
\end{equation}
These four variables, see~\cite{PRD,WHERE}, are constrained by four
relationships:\\

\noindent
(a) the luminosity equation, which tells  that  the fusion of four protons
(and two electrons) into one $\alpha$ particle is accompanied by the emission
of two neutrinos, whichever is the cycle:
\begin{equation}
\label{lum}
   K_{\odot}=
\sum_i \left( \frac{Q}{2} - \langle E \rangle _i \right ) \Phi_i
\end{equation}
where $K_{\odot}$ is the solar constant
($K_{\odot}=8.533\cdot 10^{11}$ MeV cm$^{-2}$ s$^{-1}$), Q=26.73 MeV and
$\langle E \rangle _i $ is the average energy of the i-th neutrinos.\\

\noindent
(b) The Gallium signal $S_G$=$(74\pm8)$ SNU (weighted average between the
Gallex~\cite{Gallex} and SAGE~\cite{Sage} results)
can be expressed as a  linear combination of
the $\Phi_i$'s, the weighting factors $\sigma_{i,G}$ being the absorption cross
section  for the i-th neutrinos, averaged on their energy spectrum,
see Table~\ref{sigma} for updated values:
\begin{equation}
\label{sga}
S_G= \sum_i \sigma_{i,G} \Phi_i
\end{equation}

\noindent
(c) A similar equation holds for the Chlorine
experiment, $S_C=(2.55\pm0.25)$SNU~\cite{Davis}:
\begin{equation}
\label{scl}
S_C= \sum_i \sigma_{i,C} \Phi_i
\end{equation}

\noindent
(d) The Kamiokande experiment determines - for standard neutrinos - the flux
of Boron neutrinos~\cite{Kam}:
\begin{equation}
\label{ska}
     \fib= (2.73\pm0.38)\cdot 10^6  {\mbox{cm$^{-2}$s$^{-1}$}}.
\end{equation}

With the numerical values in Table~\ref{sigma}, from Eq.~(\ref{lum})
(after dividing both term by $Q/2$) and Eqs.
(\ref{sga}) and (\ref{scl}) one gets:
\begin{eqnarray}
\label{numeq}
63.85    &=&     0.980 \fip +0.939 \fibe \nonumber\\
         &+&    0.937 \ficno +0.498 \cdot 10^{-3}\fib \nonumber\\
         & &  \nonumber \\
74\pm 8  &=&    1.23 \fip +7.32 \fibe \nonumber\\
         &+& 8.72 \ficno +2.43 \fib \nonumber \\
          & &  \nonumber \\
2.55\pm 0.25  &=& 0.38\cdot 10^{-2} \fip +0.24 \fibe \nonumber\\
         &+&   0.41 \ficno +1.11 \fib  \, ,
\end{eqnarray}
where all fluxes are in units of $10^9$~cm$^{-2}$~s$^{-1}$, but the B~flux
which is in units of $10^6$~cm$^{-2}$~s$^{-1}$.
Only errors on experimental signals are kept, since, to a first
approximation, they are dominant for determining fluxes.

The three equations~(\ref{lum}), (\ref{sga}) and (\ref{scl}) together with
(\ref{ska}) imply a unique solution:
\begin{eqnarray}
\label{solution}
\fibe  &= & (0.4\pm6.6)\cdot 10^9 {\mbox {cm$^{-2}$s$^{-1}$}} \nonumber \\
\ficno &= &(-2.0\pm 4.8)\cdot 10^9 {\mbox {cm$^{-2}$s$^{-1}$}} \nonumber \\
\fip   &= &(66.7 \pm 2.0)\cdot 10^9 {\mbox {cm$^{-2}$s$^{-1}$}} \, .
\end{eqnarray}

One notes that the central value for $\ficno$ is unphysically negative.
At first sight, this seems not to be a problem, in view of the estimated
error. However, there is a strong correlation between the errors.

In order to understand what is going on, and to  make clear the role of
each experimental result, let us reduce the number of equations and of
unknowns by the following tricks:\\
(a) one can eliminate $\fip$ by using the luminosity
    equation~(\ref{lum});\\
(b) since $\langle E\rangle _{CNO} \geq \langle E \rangle _{Be} $,
 the corresponding cross section has to be larger
than that of Be neutrinos. Thus the minimal CNO signal is obtained with
the replacement
\begin{equation}
	\sigma_{CNO}\rightarrow \sigma_{Be}
	\end{equation}
(We remark that this is also a safe approach, since the theoretical value
of $\sigma_{Be,G}$ has essentially been verified to the 10\%
level by the Gallex
neutrino source experiment~\cite{calibration}).

In this way, the above equations can be written in terms of two variables,
$\fibecno$ and $\fib$,  and the results of each experiment can be plotted in
the ($\fib$, $\fibecno$) plane, see Fig.~\ref{fig1}.

Clearly all four experiments point towards $\fibecno < 0$. This means that
the
statement ``{\em neutrinos are standard and experiments are correct}''
has lead us
to an unphysical conclusion. Could the problem be with some experiment? It
is clear from Fig.~\ref{fig1} that the situation is unchanged by
arbitrarily disregarding one of the experiments, see Ref.~\cite{wrong}.

As an attempt of being more quantitative, by applying standard statistical
arguments to Eqs.~(\ref{numeq}) we can derive the following conclusions:\\
(1) the chance $P$ for the unknown variable $\fibecno$ to be positive is less
than about 2\%. Should we disregard arbitrarily one of the experiments,
still one has $P\leq 6\%,7\%$ or $8\%$ neglecting respectively the results of
Chlorine, Gallium or Kamiokande.
This indicates that standard neutrinos ($\fibecno\geq 0$) are unlikely.\\
(2) To the 99.5\% C.L., the unknown variable $\fibecno$ should not exceed
    0.7$\cdot10^{9}$~cm$^{-2}$~s$^{-1}$.\\
(3) To the same confidence level, if one assumes {\em a priori}
standard neutrinos (and therefore $\fibecno\geq 0$) the combined flux of
Be and CNO neutrinos does not exceed 2$\cdot 10^{9}$~cm$^{-2}$~s$^{-1}$.\\
(4) Similar statements hold for Be flux (take $\ficno=0$) and CNO flux
    (put $\fibe=0$ in Eqs.~(\ref{numeq})), see Table~\ref{table1}.

The main message can be roughly summarized by saying that the chances of
standard neutrinos are low, not much more than 2\%. However, some
caution is needed, since the experimental errors we are using are
combinations of statistical fluctuations and systematic uncertainties.

\begin{table*}[htb]
\setlength{\tabcolsep}{1.4pc}
\caption[model]{
Information on Be, CNO and B neutrino fluxes.
Be and CNO (B) fluxes are in units of
$10^9$cm$^{-2}$ s$^{-1}$ ($10^6$cm$^{-2}$ s$^{-1}$).
All bounds are at the 99.5\% C.L. Direct information (a) is only available
for B neutrinos, from Ref.~\cite{Kam}. The bounds in (b) correspond to
no prior knowledge on the unknown variables. In (c) we assume {\em a
priori} $\Phi_i \geq 0$.
The results of SSMs with diffusion are also shown:
P94 from Ref.~\cite{P94}, BP95 from Ref.~\cite{BP95},
FRANEC95 indicates our preliminary results~\cite{Ciacio}.}
\label{table1}
\vspace{0.2cm}
\begin{tabular}{lcccccc}
\hline
flux &(a)&(b)&(c)&P94&BP95&FRANEC95\\
\hline
B&2.73$\pm$1.14&&&6.48&6.62&6.9\\
Be+CNO&&$\leq$0.6&$\leq$1.9&6.38&6.31&6.5\\
Be&&$\leq$0.6&$\leq$1.9&5.18&5.15&5.3\\
CNO&&$\leq$0.4&$\leq$1.4&1.20&1.16&1.2\\
\hline
\end{tabular}
\end{table*}

\section{Experimental results and standard solar models}
Let us insist on the hypothesis of standard neutrinos and compare
experimental information with theoretical estimates.

\begin{figure}[htb]
\vspace{-1.0cm}
\epsfig{file=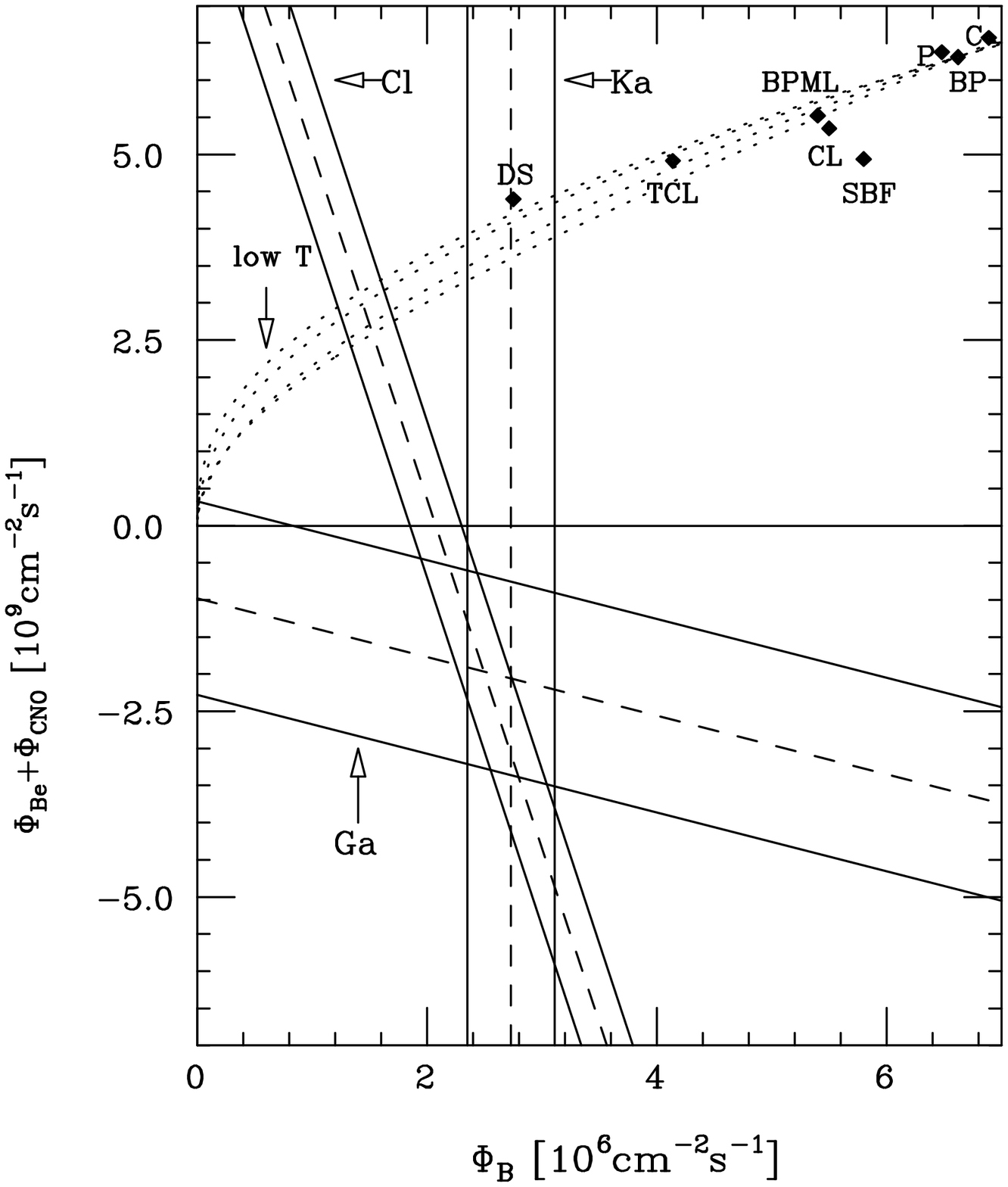,
width=1.0\hsize}%
\vspace{-2.0cm}
\caption[aaa]{Neutrino fluxes  allowed by the present experimental results.
Dashed lines correspond to central values of the experimental
results, solid lines denote $\pm 1\sigma$
limits. Diamonds represent recent solar model
calculation~\cite{DS,P94,TCL,BP95,CESAM,SAM,CL,Ciacio}.
The dotted area corresponds to (non-standard) low temperature solar
models~\cite{PRD,PLB}.
             }
 \label{fig1}
\end{figure}

We have reported in Fig.~\ref{fig1} the results of several recent solar model
calculations (diamonds) \cite{DS,P94,TCL,BP95,CESAM,SAM,CL,Ciacio}
together with experimental results.
Some of the models predict a B flux  close to the Kamiokande value; however
no model is  capable of reproducing the low Be+CNO flux implied
by the experiments.

%
In Table~\ref{table1}, we have  considered only standard solar models
where He and heavier element diffusion is taken into
account~\cite{P94,BP95,Ciacio}, as these should be more accurate.
Indeed, the comparison with helioseismology tells us
that diffusion is important for  solar models  to predict the correct
depth of the convective envelope~\cite{CD,Prof}.
We also note that in models with diffusion the central solar temperature
is increased: as Helium falls towards the centre, the mean
molecular weight increases in the stellar core and a higher temperature is
needed to balance the gravitational force.  Models with diffusion yield
thus even larger Be, CNO and B  neutrinos fluxes.

For standard neutrinos, the experimental information is presented in
columns a) and c) of Table~\ref{table1}. The discrepancy between theory and
experiment is about a factor two for the Boron flux. More important looks
to us the discrepancy on $\fibecno$, where the predicted values exceed the
experimental upper bounds by a factor three, at least.

The problem is mostly with beryllium neutrinos and let us examine
it in some detail. The extraction of $\fibe$
from experimental data (with the requirement $\ficno \geq 0$) yields an
unphysically negative Be flux. Without any prior knowledge, $\fibe$ cannot
exceed 1/10 of the SSM prediction at the 99.5\% C.L. If we {\em a priori}
force it to be non negative, the upper bound is 1/5 of the SSM at the
95\% C.L.; a value as high as 1/3 of the SSM prediction is only allowed
at the 99.5\% C.L. All this indicates that Be neutrino suppression
is much stronger than that of B neutrinos.

\section{The relevance of Beryllium}

As well known, theoretical predictions are more robust for Be than for B
neutrinos, the reasons being the weaker sensitivity to the central solar
temperature  $T$ and the independence on  the (poorly known) astrophysical
factor $S_{17}$ for the p+$^7$Be$\rightarrow ^8$B+$\gamma$ reaction.
Approximately, one has:
\begin{eqnarray}
\label{flux}
\fibe &=&\Phi_{\mbox{\footnotesize Be},0} (T/T_0)^{10}  \nonumber \\
\fib  &=&\Phi_{\mbox{\footnotesize B},0}  (S_{17}/S_{17,0}) (T/T_0)^{20}
\, ,
\end{eqnarray}
where the subscript 0 refers here and in the following to the SSM
predictions. For the power law coefficients see~\cite{PRD,libro,Bahnew}.

In addition, we point out a  relationship between $\fibe$ and $\fib$
which elucidates physically the problem of the relative abundances of Be
and B neutrinos.
\begin{figure}[htb]
\vspace{-1.0cm}
\epsfig{file=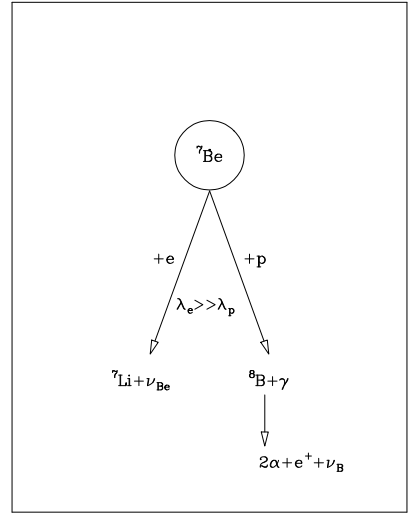,
width=1.0\hsize}%
\vspace{-2.0cm}
\caption[aaa]{The fate of $^7$Be nuclei.}
\label{fig2}
\end{figure}
Both Be and B-neutrinos  are sons of the $^7$Be nucleus, see
Fig.~\ref{fig2}. For this nucleus, electron capture (rate $\lambda_e$)
is clearly favoured over proton
capture (rate $\lambda_p$),
due to the absence of the Coulomb barrier (it is  curious that a
weak process has a larger chance than an electromagnetic process, but this
is the case due to the exponentially small penetration probabilities of
the Coulomb barrier, at the energies of interest to us). Thus  the value
of $\fibe$ is a clear indicator of
the central density $n_7$ of the
 progenitors  $^7$Be nuclei:
\begin{equation}
n_7 \propto \fibe /  \lambda_{e}
\end{equation}
If $\fibe$ comes out to be reduced by some (large) factor with respect to
the SSM prediction, the same holds for the $^7$Be equilibrium abundance
(we recall that $\lambda_e$ is weakly dependent on temperature, and it is
essentially known from measurements in the laboratory, see
Ref.~\cite{Rolfs}~).
The puzzle is thus with B neutrinos, since:
\begin{equation}
\fib \propto n_7 \lambda_{p}
\end{equation}
The observed (Kamiokande) value of $\fib$ being just a factor
two below the SSM prediction, it looks that experiments are observing too
high $\fib$! Put it in another way, one cannot kill the
father/mother  before the baby is conceived.

Should we insist on this road, we need to enhance $\lambda_{p} /
\lambda_{e}$. We
remark that any attempt to reduce $S_{17}$ goes into the wrong direction.

\section{Reduced  central temperature models? }

Non standard solar models with smaller central temperaure can be obtained
by varying -- well beyond the estimated uncertainties -- a few  parameters
(the cross section of the pp reaction,  chemical composition, opacity,
age... \cite{PRD,PLB}). These models span the dotted area in
Fig.~\ref{fig1},
which can be clearly understood by simple considerations.

To a rough approximation, also $\ficno$ has a power law dependence on
temperature~\cite{PRD,libro,Bahnew}:
\begin{equation}
\ficno=\Phi_{\mbox{\footnotesize CNO},0} (T/T_0)^{20}
\end{equation}
One can use this equation together with Eqs.~(\ref{flux}) above; by
expressing the temperature as a function of $\fib$, one has:
\begin{eqnarray}
\fibe + \ficno &=&
\Phi_{\mbox{\footnotesize Be},0} (\fib /
 \Phi_{\mbox{\footnotesize B},0})^{1/2} \nonumber \\
             &+&
\Phi_{\mbox{\footnotesize CNO},0} (\fib / \Phi_{\mbox{\footnotesize B},0})
\, ,
\end{eqnarray}
and one sees in Fig.~\ref{fig1} the square root behaviour at small $\fib$,
 which then
changes to linear for larger $\fib$.

It is clear that all these model fail  to reproduce the experimental
results, essentially because they cannot reproduce the observed ratio
$\fibe / \fib $, see Ref.~\cite{Bere}, as a consequence of the drastically
different dependences on temperature, see Eqs.~(\ref{flux}).
If $\fib$ is reduced by a factor two, $\fibe$ is too high. On the other hand,
if $\fibe$ is brought to the low level required by the experiments, the
predicted $\fib$ is definitely too small. In other words, as we said
previously,  we are observing too many B-neutrinos (if neutrinos are
standard)!
\begin{table*}[htb]
\setlength{\tabcolsep}{1.4pc}
\catcode`?=\active \def?{\kern\digitwidth}
\caption[borex]{
Predictions for Beryllium neutrinos. For different models
we present at the best fit point (sin$^2 2\theta$, $\delta m^2$), the
$\chi^2$ for degree of freedom, the flux and signal (CC+NC) in units of
the SSM predictions.
               }
\label{table2}
\vspace{0.2cm}
\begin{tabular}{llccccc}
\hline
&&$\chi^2/d.o.f.$&$\sin^2 2\theta$&$\delta m^2$[eV$^2$]&$\Phi/\Phi_0$&S/S$_0$\\
 \hline
(a) & \multicolumn{6}{l}{Active neutrinos:}\\
&MSW small $\theta$&0.9/2&0.0058&7.9$\cdot10^{-6}$&9\%&27\% \\
&MSW large $\theta$&1.5/2&0.63&1.7$\cdot10^{-5}$&48\%&59\%\\
&Just-So	&1.9/2&1.00&6.0$\cdot10^{-11}$&78\%&82\%\\
\hline
(b)& \multicolumn{6}{l}{Sterile neutrinos:}\\
&MSW small $\theta$&0.7/2&0.0079&4.9$\cdot10^{-6}$&2\%&2\%   \\
&MSW large $\theta$&8.1/2&0.73&1.3$\cdot10^{-5}$&46\%&46\% \\
&Just-So	&7.2/2&0.86&6.2$\cdot10^{-11}$&33\%&33\% \\
\hline
\end{tabular}
\end{table*}

\section{Higher central temperatures? (Or why do we care about CNO
neutrinos)}

One could imagine the conspiracy of two mechanisms, so as to bring both
$\fibe$ and $\fib$ in agreement with experiment. For example, one could assume
that $S_{33}$ is much larger than commonly assumed (e.g. as a result of a
hypothetical resonance~\cite{Clay}) so as to enhance the ppI channel and
reduce $\fibe$ to the desired value. At the same time, by varying some
suitable  parameter the central temperature could be increased, so as to
bring $\fib$ in agreement with experiment.

\begin{figure}[htb]
\vspace{-1.0cm}
\epsfig{file=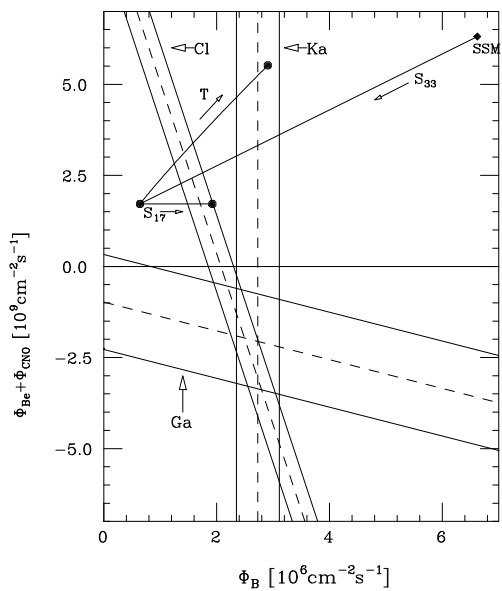,
width=1.0\hsize}%
\vspace{-2.0cm}
\caption[lmts23ex]{Sketch  of the behaviour of solar models
with non standard $S_{33}$, central temperature T and $S_{17}$,
from Ref.~\cite{last}.
                  }
\label{fig3}
\end{figure}

This mechanism also fails~\cite{last}, see Fig.~\ref{fig3},
due to
the fact that as temperature raises, the CNO flux grows as fast as the
Boron flux, and the experimental bound on $\fibecno$ is again violated.

In other words, while Beryllium and Boron neutrinos tell us that one cannot
hope to solve the neutrino problem by lowering the central temperature,
the bound on CNO implies that  increasing the temperature does not work
either.

If instead the temperature is unchanged but $S_{17}$ is increased, one
has still the problem that the  SSM prediction for the CNO exceeds the
experimental constraint, see again Fig.~\ref{fig3} and Ref.~\cite{last}.

\section{Expectations for Be  neutrinos}

We have seen that, for standard neutrinos, the Be-flux is strongly
suppressed with respect to the SSM predictions. What has to be expected for
non standard neutrinos?

In Table~\ref{table2} we update and extend a recent analysis~\cite{justso}
for a few candidate solutions. We use now as a reference the fluxes
corresponding to the ``best model with Helium and metal diffusion'' of
Ref.~\cite{BP95}. For active
neutrinos, both  small and large angle  MSW solutions are acceptable, as
well as the Just-So model. On the other hand, for sterile neutrinos only
MSW at small angle gives a good fit. Among these four acceptable solutions,
two of them (MSW large angle and Just-So) give signals (CC+NC) that are
quite a significant fraction of the SSM prediction, see last column in
Table~\ref{table2}.
In other words, in face of the present experimental data, the
Beryllium signal  does not need to be small, for non standard neutrinos.

\begin{figure}[htb]
\vspace{-1.0cm}
\epsfig{file=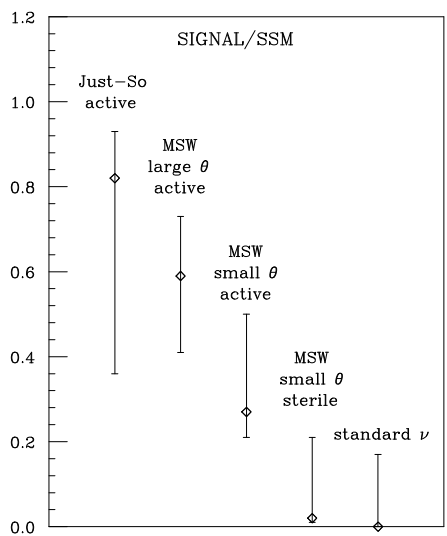,
width=1.0\hsize}%
\vspace{-2.0cm}
\caption[lmts23e]{The Beryllium (CC+NC) signal, in units of the
SSM prediction. Diamonds indicate the best fit points, bars correspond
to 90\% C.L.
                 }
\label{fig4}
\end{figure}

The situation is made more clear in Fig.~\ref{fig4}, where we show
the 90\%~C.L. regions  according to the different models.
A direct measurement of the Be line can in many cases  discriminate
among the possibile solutions. Very large signals, above 75\% of the SSM
prediction, correspond essentially to the Just-So solution. Between 75\%
and about 35\% various models are acceptable. Between 35\% and 20\% the
solution has to bee MSW at small angle for active neutrinos.  Very small
signal, say below 20\%, are only possible for standard neutrinos, or
transitions into sterile neutrinos.

In the intermediate region discrimination between Just-So and MSW solutions
should be obtained by Borexino looking at seasonal variations, even for
purities well below the  design  purity, see~\cite{justso}.

\begin{center} *** \end{center}

Most of the results presented here are the outcome of the fruitful and
friendly collaboration in the last few years with V.~Berezinsky,
V.~Castellani, F.~Ciacio, S.~Degl'Innocenti and N.~Ferrari.
We are grateful to T.~E.~O.~Ericson and to G.~Cocconi for useful
discussions. One of us (G.~F.) thanks the CERN Theory Division for
hospitality while this work was done.
\end{document}